\documentclass[]{aa}  
\usepackage{graphicx}
\usepackage{graphics}
\usepackage{txfonts}
\usepackage{natbib}
\usepackage[retainorgcmds]{IEEEtrantools}
\usepackage{rotating}
\usepackage{subfigure}
\usepackage{caption}

\bibpunct{(}{)}{;}{a}{}{,} 

\begin{document}
   \title{Modulated cycles in an illustrative solar dynamo model with competing $\alpha$-effects}
   \titlerunning{Competing $\alpha$-effects in a solar dynamo model}
   \author{L.C. Cole
          \and
          P.J. Bushby
}
   \institute{School of Mathematics and Statistics, Newcastle University,
              Newcastle Upon Tyne, NE1 7RU, UK\\
              \email{l.cole@ncl.ac.uk,paul.bushby@ncl.ac.uk}
 }  
   
 
  \abstract
{The large-scale magnetic field in the Sun varies with a period of approximately 22 years, although the amplitude of the cycle is subject to long-term modulation with recurrent phases of significantly reduced magnetic activity. It is believed that a hydromagnetic dynamo is responsible for producing this large-scale field, although this dynamo process is not well understood.}
{Within the framework of mean-field dynamo theory, our aim is to investigate how competing mechanisms for poloidal field regeneration (namely a time delayed Babcock-Leighton surface $\alpha$-effect and an interface-type $\alpha$-effect), can lead to the modulation of magnetic activity in a deep-seated solar dynamo model.}
{We solve the standard $\alpha\Omega$ dynamo equations in one spatial dimension, including source terms corresponding to both of the the competing $\alpha$-effects in the evolution equation for the poloidal field. This system is solved using two different methods. In addition to solving the one-dimensional partial differential equations directly, using numerical techniques, we also use a local approximation to reduce the governing equations to a set of coupled ordinary differential equations (ODEs), which are studied using a combination of analytical and numerical methods.}
{In the ODE model, it is straightforward to find parameters such that a series of bifurcations can be identified as the time delay is increased, with the dynamo transitioning from periodic states to chaotic states via multiply periodic solutions. Similar transitions can be observed in the full model, with the chaotically modulated solutions exhibiting solar-like behaviour.}
{Competing $\alpha$-effects could explain the observed modulation in the solar cycle.}

\maketitle
%
\section{Introduction}

At the solar photosphere, bipolar active regions are formed when loops of magnetic flux rise to the surface from the base of the convection zone due to the action of magnetic buoyancy \citep[][]{PARKER3}. This implies that the properties of sunspot-bearing active regions can be used to deduce some of the features of the underlying large-scale magnetic field. It is well known \citep[see, for example,][]{STIX,CHARB,JONES} that zones of active region emergence follow a cyclic pattern with a period of approximately 11 years. At the beginning of each cycle, sunspots tend to be found at mid-latitudes, with zones of emergence drifting towards the equator as the cycle progresses. The underlying large-scale (predominantly azimuthal) magnetic field changes sign at the end of each cycle, giving a full magnetic period of approximately 22 years. However, the solar cycle is not strictly periodic. In particular, the peak amplitude (measured, for example, by the sunspot coverage) varies from one cycle to the next. Although this modulation does not usually disrupt the cycle, more extreme episodes of modulation have been recorded. For example, during a period known as the Maunder Minimum, very few sunspots were observed between approximately 1650 and 1720  \citep[][]{EDDY,RIBES}. However, sunspot records are not the only indicators of modulation. Due to the fact that the Sun's strong magnetic field protects the Earth from cosmic rays, the abundance of certain isotopes in the Earth's atmosphere is known to be anti-correlated with the solar cycle. Therefore, by analysing Beryllium-10 deposits in ice cores \citep[see, for example,][]{BE10} and Carbon-14 levels in tree rings \citep[see, for example,][]{C14} it is possible to deduce the history of the solar cycle. Such studies have indicated that cyclic activity did persist throughout the Maunder Minimum, but at a significantly reduced level \citep[][]{BEER}. Furthermore, it is clear that the Maunder Minimum is not exceptional -- the solar cycle has often been interrupted by recurrent ``Grand Minimum'' phases of significantly reduced magnetic activity.

\par It is believed that the large-scale magnetic field in the solar interior is generated and maintained by a hydromagnetic dynamo. From a conceptual point of view, the large-scale field can usefully be decomposed into its toroidal (azimuthal) and poloidal (meridional) components -- a working dynamo requires mechanisms that allow the poloidal field to be regenerated from toroidal field and vice versa. It is widely accepted that differential rotation (usually referred to as the $\Omega$-effect in dynamo theory) is responsible for the generation of toroidal field from poloidal field. Surface observations indicate that equatorial regions rotate more rapidly than the poles, and helioseismological studies \citep[see, for example,][]{SCHOU} have shown that this rotation profile persists, approximately independently of radius, throughout most of the convection zone. At the base of the convection zone, a region of strong shear (the tachocline) couples the radiative zone, which rotates almost rigidly, to the differentially-rotating convective envelope. In most solar dynamo models, it is assumed that a significant fraction of the toroidal field is generated in the vicinity of the tachocline (where the $\Omega$-effect should be very efficient due to the presence of strong differential rotation).   

\par Although the $\Omega$-effect is well understood, the reverse process that generates poloidal field from toroidal field is a topic of some debate. In classical interface dynamo models \citep[see, for example,][]{PARKER,CHARBMAC} the poloidal field is regenerated at the base of the convection zone by the action of cyclonic convection upon toroidal magnetic field lines \citep[][]{PARKER2}. This process is usually referred to as the $\alpha$-effect. Strong toroidal fields will tend to inhibit (or quench) the operation of the $\alpha$-effect, so interface dynamo models are usually constructed in such a way that the $\alpha$-effect is restricted to the region just above the base of the convection zone, whilst the $\Omega$-effect operates just below the interface. The two layers are coupled by the effects of magnetic diffusion, as well as magnetic buoyancy and turbulent pumping \citep[see, for example,][]{TOBBRU}. Even with strong $\alpha$-quenching, it has been shown that an interface dynamo of this type can operate efficiently \citep[][]{CHARBMAC}.  In Babcock-Leighton dynamo models \citep[][]{BL,LTON1}, the poloidal field is regenerated at the solar surface through the decay of active regions (which tend to emerge with a systematic tilt with respect to the azimuthal direction). This surface $\alpha$-effect can only contribute to the dynamo if there is some mechanism that is capable of transporting the resultant poloidal field to the tachocline. This could be achieved by diffusion or by pumping, but meridional flows also could play an important role in this respect. A polewards meridional flow is observed at the solar surface \citep[see, for example,][]{HATHAWAY} and, by mass conservation arguments, there must be a returning circulatory flow somewhere within the solar interior. A single-cell meridional circulation, with an equatorial flow at the base of the convection zone would couple the surface layers to the tachocline in an effective way, thus completing the dynamo loop. 

\par A complete model of the solar dynamo must be able to explain the observed modulation as well as the 22-year magnetic cycle. It has been shown that it is possible to induce modulation by introducing stochastic effects into Babcock-Leighton models \citep[][]{DIKCHARB,BUSHTOB}, as well as into models of interface type \citep[][]{OSSENDRIJVER2000}. However, fully deterministic models (with no random elements) can also produce modulated dynamo waves. \citet{WEISSCATJONES} and \citet{JWC} considered a simple system in which the dynamo was modelled using a set of coupled ordinary differential equations, which included the nonlinear interactions between the magnetic field and the flow. They found that it was possible to generate quasiperiodic and chaotically-modulated solutions in addition to standard periodic dynamo waves. More recent studies have shown that the full mean-field equations also exhibit significant modulation when dynamical nonlinearities are included in the governing equations \citep[see, for example,][]{TOBIAS96,BROOKE,BUSHBY}.  An alternative approach was used by \cite{YOSHI} who demonstrated that modulation can arise if explicit time delays are built into the nonlinear terms in a simple system of model dynamo equations. A more sophisticated model was considered by \cite{JOUVE} who investigated the effects of magnetic buoyancy-induced time delays in the context of a two-dimensional Babcock-Leighton dynamo. By introducing time delays into the surface $\alpha$-effect term, they were able to demonstrate the existence of modulated cycles. They then went on to consider a simpler one-dimensional dynamo system in which the surface $\alpha$-effect term was represented by the inclusion of a time-delayed toroidal field (with a parameterised time delay that was dependent upon the magnetic field strength). They were able to demonstrate the existence of a sequence of bifurcations from periodic to chaotically modulated solutions as the time delay parameter was increased.  

\par The aim of this work is to investigate the competition between a deep-seated (interface) $\alpha$-effect and a surface $\alpha$-effect. Building on the approach described by \citet{JOUVE}, who did not include a deep-seated $\alpha$-effect, the influence of the surface $\alpha$-effect will be modelled using a time-delayed toroidal field. The use of a time delay is natural in this context: even if flux tubes rise rapidly  to the surface, the time taken for the resultant poloidal field to be transported back to the tachocline will, in general, be non-negligible compared to the period of oscillation of the dynamo. Previous studies have investigated systems with competing $\alpha$-effects \citep[see, for example,][]{DIKPATI01,MASON,MANN}, but we believe that this is the first study to consider the effects of explicit time delays in a model of this type. The paper is structured as follows: In Section 2, we describe the full model and an idealised system of equations that can derived from it (based upon a local analysis). This is followed in Section 3 by an analysis of the stability of the idealised model and then in Section 4 by the corresponding numerical results. In Section 5, we describe some numerical calculations which demonstrate the existence of modulated solutions in the full one-dimensional model. Finally, in Section 6, we present our conclusions and discuss the relevance of our results to the solar dynamo.

\section{\label{sec:compalpha}Model Setup}

Following a similar approach to that adopted by \citet{JOUVE}, we consider a simple, illustrative model of the solar dynamo. This model is based upon the standard mean-field dynamo equation \citep[see, e.g.,][]{MOFFATT},

\begin{equation}
\frac{\partial \vec{B}}{\partial t} = \nabla \times \left(\alpha \vec{B} + \vec{U}\times\vec{B}\right) + \eta_T \nabla^2 \vec{B},\label{MF}
\end{equation}
\noindent where $\vec{U}$ is the large-scale velocity field, $\alpha$ represents the standard mean-field $\alpha$-effect, $\eta_T$ is the turbulent magnetic diffusivity (which we shall assume to be constant), whilst the mean magnetic field, $\vec{B}$, satisfies $\nabla\cdot\vec{B}=0$.  Instead of solving this equation in spherical geometry, we consider the simpler problem of dynamo action in a flat Cartesian domain, with the axes oriented so that the $y$-axis would correspond to the azimuthal direction on a spherical surface. We can then look for dynamo solutions that depend only on a single spatial variable $x$ (which can be regarded as being analogous to the co-latitude) and time $t$. The solenoidal constraint upon $\vec{B}$ can then be satisfied by writing the magnetic field in the following form:

\begin{equation}  
\vec{B}(x,t)=B(x,t)\hat{\vec{y}}+\nabla\times\left[A(x,t)\hat{\vec{y}}\right],\label{POLTOR}
\end{equation}
\noindent where $B(x,t)$ is the toroidal field component, whilst $A(x,t)$ corresponds to the poloidal potential. 

\par Our model is based on the assumption that the solar dynamo is operating primarily in the region around the base of the convection zone. For simplicity, we assume that $\alpha$, which represents a deep-seated $\alpha$-effect, is constant, i.e. $\alpha=\alpha_0$. Furthermore, we adopt a fixed velocity profile of the form $\vec{U}=v_0\hat{\vec{x}}+\Omega_0 z \hat{\vec{y}}$, where $v_0$ and $\Omega_0$ are both assumed to be constant in this illustrative model. This velocity field gives a constant meridional flow and a differential rotation profile that is independent of $x$. We also make the well known $\alpha\Omega$ approximation, which assumes that differential rotation is the dominant mechanism for toroidal field regeneration in this region. Following \citet{JOUVE}, we also introduce a delayed toroidal field, $Q(x,t)$, which lags behind the normal toroidal field with a time delay denoted by $\tau$. However, unlike \citet{JOUVE}, who considered a time delay that was dependent upon the toroidal magnetic field strength, we assume $\tau$ to be constant throughout this study. The delayed toroidal field is coupled to the other equations via the inclusion of an additional poloidal source term, $SQ(x,t)$, where $S$ is a constant. This source term can be regarded as being the contribution to the local poloidal field from the non-local surface $\alpha$-effect (which must, therefore, depend upon the strength of the toroidal field at earlier times). Finally, we introduce parameterised quenching nonlinearities into both of the $\alpha$-effect terms in the poloidal field equation. 

\par Having made these assumptions, we can now write down the three scalar partial differential equations for $A(x,t)$, $B(x,t)$ and $Q(x,t)$:

 \begin{IEEEeqnarray}{rCl}
 \frac{\partial A}{\partial t} &+& v_0\frac{\partial A}{\partial x} = \frac{SQ}{1+\lambda|Q|^2}+\frac{\alpha_0 B}{1+\lambda|B|^2}+\eta_T\frac{\partial^2A}{\partial x^2},\label{PDEfinala}\\
 \frac{\partial B}{\partial t} &+& v_0\frac{\partial B}{\partial x} = \Omega_0\frac{\partial A}{\partial x}+\eta_T\frac{\partial^2B}{\partial x^2},\label{PDEfinalb}\\
 \frac{\partial Q}{\partial t} &=& \frac{1}{\tau}\left(B-Q\right),\label{PDEfinalq}
\end{IEEEeqnarray}
\noindent where $\lambda$ is a constant that determines the strength of the nonlinear quenching. In order to reduce the number of parameters that control the system, the variables can be rescaled as follows:

 \begin{IEEEeqnarray*}{rCl}
 A&=&\frac{\alpha_0B_0L^2}{\eta_T}A',~~~~~B=B_0B',~~~~~Q=B_0Q',~~~~~t=\frac{L^2}{\eta_T}t',\\
 \tau&=&\frac{L^2}{\eta_T}\tau',~~~~~~~~~~x=Lx',~~~~~~~~~~S=\alpha_0S',
\end{IEEEeqnarray*}
\noindent where $L$ is a characteristic length-scale and $B_0$ is a representative value of the magnetic field strength (which may be chosen so that the constant coefficient in the quenching terms equals unity in these scaled variables). On dropping the primes, we obtain the following set of partial differential equations (PDEs):  

 \begin{IEEEeqnarray}{rCl}
 \frac{\partial A}{\partial t} &+& Re\frac{\partial A}{\partial x} = \frac{SQ}{1+|Q|^2}+\frac{B}{1+|B|^2}+\frac{\partial^2A}{\partial x^2},\label{PDEa}\\
 \frac{\partial B}{\partial t} &+& Re\frac{\partial B}{\partial x} = D\frac{\partial A}{\partial x}+\frac{\partial^2B}{\partial x^2},\label{PDEb}\\
 \frac{\partial Q}{\partial t} &=& \frac{1}{\tau}\left(B-Q\right).\label{PDEq}
\end{IEEEeqnarray}

\noindent Thus the only parameters to control the system are the Reynolds number, $Re=v_0L/\eta_T$, which measures the strength of the meridional flow, and the dynamo number, $D=\alpha_0\Omega_0L^3/\eta_T^{2}$, which indicates the strength of the dynamo sources relative to magnetic dissipation. 

\par This system can be further simplified by carrying out a local analysis. Because we have the freedom to choose a convenient characteristic length-scale, local wavelike solutions can be assumed to have a unit wavenumber without any loss of generality. We therefore seek solutions of the form $A=\tilde{A}(t)e^{ix}$, $B=\tilde{B}(t)e^{ix}$ and $Q=\tilde{Q}(t)e^{ix}$, where $\tilde{A}(t)$, $\tilde{B}(t)$ and $\tilde{Q}(t)$ are complex functions of time only. Dropping the tildes, the governing equations for these quantities become: 

\begin{IEEEeqnarray}{rCl}
 \frac{dA}{dt} &+& iReA = \frac{SQ}{1+|Q|^2}+\frac{B}{1+|B|^2}-A, \label{ODEa}\\
 \frac{dB}{dt} &+& iReB = iDA-B, \label{ODEb}\\
 \frac{dQ}{dt} &=& \frac{1}{\tau}\left(B-Q\right). \label{ODEq}
\end{IEEEeqnarray}

\noindent Following the methods used in \cite{JWC} and \cite{JOUVE} it is possible to reduce the order of this system by using the following representation:

\begin{IEEEeqnarray*}{rCl}
 A &=& \rho y e^{i\theta},\\
 B &=& \rho e^{i\theta},\\
 Q &=& \rho z e^{i\theta},
\end{IEEEeqnarray*}

\noindent where $\rho$ and $\theta$ are real quantities and $y$ and $z$ are complex numbers. Upon substituting these expressions into the governing equations (\ref{ODEa}) - (\ref{ODEq}), the following set of 5 real ODEs is obtained:

\begin{IEEEeqnarray}{rCl}
 \frac{d\rho}{dt} &=& -D\rho y_2-\rho, \label{5thrhomine}\\
 \frac{dy_1}{dt} &=& \frac{Sz_1}{1+\rho^2(z_1^2+z_2^2)}+\frac{\alpha}{1+\rho^2}+2 y_1 y_2D,\label{5thy1mine}\\
 \frac{dy_2}{dt} &=& \frac{Sz_2}{1+\rho^2(z_1^2+z_2^2)}+D y_2^2-D y_1^2, \label{5thy2mine}\\
 \frac{dz_1}{dt} &=& \frac{1-z_1}{\tau}+D y_1 z_2+D y_2 z_1 -Re z_2 + z_1, \label{5thz1mine}\\
 \frac{dz_2}{dt} &=& \frac{-z_2}{\tau}-D y_1z_1+D y_2 z_2 +Re z_1 + z_2, \label{5thz2mine}
\end{IEEEeqnarray}

\noindent where $y_1$ and $y_2$ represent the real and imaginary parts of $y$ respectively and $z_1$ and $z_2$ represent the corresponding real and imaginary parts of $z$. This fifth-order system is a useful alternative representation of the local model. 


\section{\label{dt}Dynamo Transitions}

In this section, we focus upon the local model that is described by Equations (\ref{ODEa}) -- (\ref{ODEq}). To further our understanding of this system, we have carried out a series of calculations to determine the critical value of the dynamo number as the parameters $S$ and $\tau$ are varied (at fixed $Re$). It is also possible to identify the value of $\tau$ that leads to quasi-periodic solutions analytically by studying the stability of the periodic solution.

\subsection{\label{lt}Linear Theory}
The critical dynamo numbers can be calculated by linearising the governing equations (\ref{ODEa}) -- (\ref{ODEq}) and writing $A$, $B$ and $Q$ in the following form: $A=\hat{A} e^{\sigma t},B=\hat{B} e^{\sigma t}$ and $Q=\hat{Q} e^{\sigma t}$. The following characteristic equation is generated:

\begin{equation}
\left(\sigma+iRe+1\right)^{2}\left(\sigma+\frac{1}{\tau}\right)-iD\left(\sigma+\frac{1}{\tau}\right)-\frac{iSD}{\tau}=0.\label{char1}
\end{equation}

\noindent Setting the real part of the growth rate to be zero and solving the characteristic equation for the imaginary part of $\sigma$ will determine the critical value of the dynamo number, $D_c$, at which the trivial (non-magnetic) solution loses stability to oscillatory dynamo waves. Letting $\sigma=i\omega$, we obtain the following:

\begin{IEEEeqnarray}{rCl}
&-&i\omega^{3}-\frac{\omega^{2}}{\tau}-2i\omega^{2}Re-\frac{2Re\omega}{\tau}-2\omega^{2}+\frac{2i\omega}{\tau}-iRe^{2}\omega-\frac{Re^{2}}{\tau} \nonumber\\
&-&2\omega Re+\frac{2iRe}{\tau}+i\omega+\frac{1}{\tau}+D\omega-\frac{iD}{\tau}-\frac{iSD}{\tau}=0.
\label{char}
\end{IEEEeqnarray}

\noindent In the case of $S=0$ the system corresponds to a standard $\alpha\Omega$ dynamo. It is then straightforward to show that the critical dynamo number is $2$, regardless of the magnitudes of $\tau$ or $Re$. For $S\neq0$, this equation can be solved numerically using a Newton-Raphson algorithm.

\begin{figure}
\begin{center}
\resizebox{\hsize}{!}{\includegraphics[angle=270]{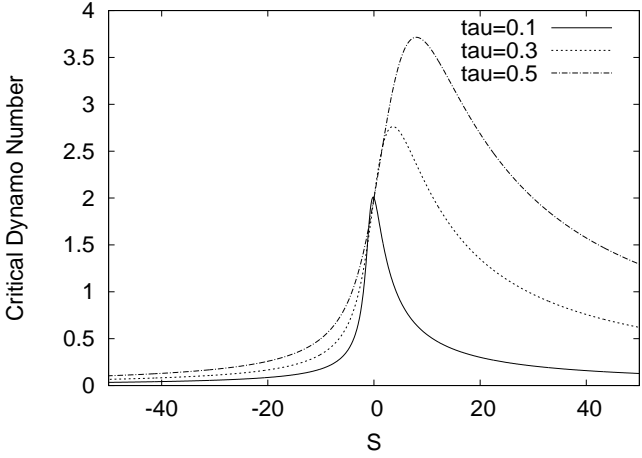}}
\caption{\label{fig:r=10t=0102030405}{The critical dynamo number $D_c$ as a function of $S$, for values of $\tau$ less than 1 with $Re=10$. Here, the solid line corresponds to $\tau=0.1$, the dotted line represents $\tau=0.3$ and the dash-dotted line shows $\tau=0.5$.}}
\end{center}
\end{figure}

As a specific example, Figure~\ref{fig:r=10t=0102030405} shows how the critical dynamo number changes as $S$ is varied between $-50$ and $50$ with values of $\tau$ less than 1 and for fixed $Re=10$. For these values of $\tau$, $D_c<2$ for most values of $S$. In these regions of parameter space, it is easier to excite a dynamo than it would be in the corresponding $\alpha\Omega$ system, so we can see that the non-local $\alpha$-effect is enhancing the dynamo for these parameter values. However, for larger values of $\tau$ (where there is a significant time-lag between $B$ and $Q$) there is a finite range of values of $S$ in which $D_c>2$, with a local maximum in $D_c$ occurring somewhere in this range. In this case, the two competing $\alpha$-effects appear to be impeding each other, thus making it more difficult to excite a dynamo. So it is clear that, even in linear theory, the interaction between competing $\alpha$-effects is non-trivial. Additional calculations have been made to determine the critical dynamo number for different values of $Re$ and a broader range of values for $\tau$, but this case (which is of greatest relevance to the present study) is fairly representative in terms of the behaviour that is exhibited as $S$ is varied. 


\subsection{\label{transition}The transition from periodic to quasi-periodic behaviour}

Restoring the nonlinear terms to the governing equations, finite amplitude oscillations can be found when the dynamo number exceeds $D_c$. The stability of the periodic solution can be analysed by expressing the magnetic fields in the following form:

\begin{IEEEeqnarray*}{rCl}
 A_{\omega}(t) &=& A_0e^{i\omega t},\\
 B_{\omega}(t) &=& B_0e^{i\omega t},\\
 Q_{\omega}(t) &=& Q_0e^{i\omega t},
\end{IEEEeqnarray*}

\noindent where $A_0$, $B_0$ and $Q_0$ are the complex wave amplitudes, $\omega$ is the frequency (with the $\omega$ subscript denoting the periodic state). The substitution of these expressions into the governing equations (\ref{ODEa}) -- (\ref{ODEq}) produces the following simultaneous equations:

 \begin{IEEEeqnarray}{rCl}
 (i\omega+iRe+1)A_0 &=& \frac{SQ_0}{1+|Q_0|^2}+ \frac{B_0}{1+|B_0|^2}\label{A_0},\\
 (i\omega+iRe+1)B_0 &=& iDA_0\label{B_0},\\
 \left(i\omega+\frac{1}{\tau}\right)Q_0 &=& \frac{1}{\tau}B_0\label{Q_0},
\end{IEEEeqnarray}
\noindent which can be solved using standard methods. Once the amplitude and frequency of the periodic solution have been determined, it is possible to perturb this solution to study its stability. Following the general method described by \citet{JOUVE}, this can be achieved by setting:

 \begin{IEEEeqnarray*}{rCl}
   &A& = A_{\omega}\left(1+\alpha_1e^{pt}+\alpha_2^\star e^{p^\star t}\right), \label{alpha}\\
   &B& = B_{\omega}\left(1+\beta_1e^{pt}+\beta_2^\star e^{p^\star t}\right), \label{beta}\\
   &Q& = Q_{\omega}\left(1+\gamma_1e^{pt}+\gamma_2^\star e^{p^\star t}\right), \label{gamma} 
 \end{IEEEeqnarray*}

\noindent where $\alpha_1,\alpha_2,\beta_1,\beta_2,\gamma_1$ and $\gamma_2$ are the coefficients of the perturbed fields, $p$ is the complex growth rate of the perturbation and the symbol $\star$ represents the complex conjugate. Substituting these expressions into the governing equations (\ref{ODEa}) -- (\ref{ODEq}) results in a system of 6 coupled equations that relates the coefficients of the perturbed fields to the growth rate $p$ for a given set of parameters. After solving this system to find the growth rate of the perturbation, it is then possible to determine the stability of the periodic solutions. 

\begin{table}
\caption{Critical values of $\tau$ for $D=1000$.}
\begin{center}
\resizebox{9cm}{!} {
\begin{tabular}{cccccccc}
\hline

~& ~& \multicolumn{6}{c}{$Re$} \\
~&~& 0 & 10 & 20 & 30 & 40 & 50 \\
\hline
~&0&(...)&(...)&(...)&(...)&(...)&(...)\\
~&10&(...)&0.348&0.271&0.259&0.260&0.268\\
$S$&30&(...)&0.334&0.255&0.239&0.226&0.208\\
~&50&(...)&0.332&0.252&0.236&0.221&0.202\\
~&70&(...)&0.330&0.250&0.234&0.218&0.200\\
\hline
\end{tabular}}
\end{center}
\caption*{The Reynolds number varies between $0$ and $50$, with $S$ varying between $0$ and $70$.}
\label{tab:tcvaluesdpos}
\end{table}

\begin{table}
\caption{Critical values of $\tau$ for $D=-1000$.}
\begin{center}
\resizebox{9cm}{!} {
\begin{tabular}{cccccccc}
\hline
~& ~& \multicolumn{6}{c}{$Re$} \\
~&~& 0 & 10 & 20 & 30 & 40 & 50 \\
\hline
 ~&0&(...)&(...)&(...)&(...)&(...)&(...)\\
 ~&-10&(...)&0.310&0.227&0.211&0.206&0.206\\
 $S$&-30&(...)&0.321&0.240&0.224&0.219&0.217\\
 ~&-50&(...)&0.323&0.243&0.226&0.221&0.219\\
 ~&-70&(...)&0.324&0.244&0.227&0.222&0.220\\
\hline
\end{tabular}}
\end{center}
\caption*{The Reynolds number varies between $0$ and $50$, with $S$ varying between $0$ and $-70$.}
\label{tab:tcvaluesdneg}
\end{table}

\begin{table*}
\caption{Values of $|B_0|$, $|Q_0|$ and $\omega$ from both analytical calculations and the numerical simulations. }
\begin{center}
\begin{tabular}{ccccccc}
\hline
~& \multicolumn{2}{c}{$Re=0$} & \multicolumn{2}{c}{$Re=10$} & \multicolumn{2}{c}{$Re=20$} \\
~& Analytic & Numerical & Analytic & Numerical & Analytic & Numerical \\
~& Calculation & Simulation & Calculation & Simulation & Calculation & Simulation \\
\hline
$|B_0|$ & 77.32 & 77.32 & 52.85 & 52.84 & 40.59 & 40.58 \\
$|Q_0|$ & 76.99 & 76.99 & 41.20 & 41.20 & 20.88 & 20.88 \\
$\omega$ & 0.92 & 0.93 & 8.03 & 8.04 & 16.67 & 16.68 \\
\hline
\end{tabular}
\end{center}
\caption*{Here, $D=1000$, $S=10$ and $\tau=0.1$, whilst the Reynolds number is varied. All results are accurate to at least $1\%$.}
\label{tab:template}
\end{table*}

Tables~\ref{tab:tcvaluesdpos} and~\ref{tab:tcvaluesdneg} illustrate some of the results from this stability analysis. These tables show the parametric dependence of the critical value of $\tau$ for the transition from periodic to quasi-periodic solutions. The results in Table~\ref{tab:tcvaluesdpos} correspond to $D=1000$, with $0\le Re\le 50$ and $0 \le S \le 70$. In Table~\ref{tab:tcvaluesdneg}, we have used the same values of $Re$, but $D=-1000$, whilst $0 \ge S \ge -70$. An entry of (...) in either table indicates that no transition exists. Unsurprisingly, no modulation is found for $S=0$. In this case the delayed toroidal field $Q$ decouples from the system and we have a standard $\alpha\Omega$ dynamo model. More unexpectedly, these results suggest that $Re \neq 0$ is a necessary condition for modulation in this system. So the meridional flow seems to play a crucial role in driving the modulation, perhaps by introducing an additional (advective) time-scale into the problem. {If it is simply the presence of an additional time-scale that is the key ingredient here, then it may still be possible to drive modulation in the absence of a flow if some other physical process (such as turbulent pumping) was included in the model. However, it is beyond the scope of this paper to investigate whether or not this is indeed the case.} In the case of positive $D$, no modulation was found for negative $S$, whilst the same is true for positive values of $S$ in the negative $D$ case. Given the idealised nature of this local model, we should probably not read too much into this result, but (if nothing else) this again illustrates that competing $\alpha$-effects interact in a rather non-trivial way in this system. Where modulation does occur, some trends can be identified. For example, for fixed $Re$ in the $D=1000$ case, the critical value of $\tau$ decreases with increasing $S$ (whereas it increases with increasing $|S|$ in the $D=-1000$ case). At fixed $S$, the critical value of $\tau$ tends to decrease with increasing values of $Re$, although this trend appears to reverse at low $S$ and high $Re$ in the $D=1000$ case. We have no definitive physical explanation for this behaviour but can speculate that this is somehow related to the non-monotonicity that was observed in the $D_c$ calculations in the previous subsection. 


\section{Numerical simulations of the local model}

In this section, we apply a numerical approach to the local model that is described by equations (\ref{ODEa}) -- (\ref{ODEq}). Decomposing the system into its real and imaginary parts, we use a fourth-order Runge-Kutta scheme in Fortran to time-step the governing equations.  

\subsection{\label{vac}Validation of numerical calculations}

To validate the code, it is possible to check that the results agree with the critical dynamo numbers that can be obtained from linear theory (see Section~\ref{lt}). Fixing the values of $Re=10$, $S=20$ and $\tau=10$, gives a prediction of $D_c=2.499$. This is consistent with the numerics: we find decaying oscillations for $D=2.4$, whilst $D=2.6$ gives a stable periodic solution. We can also compare the amplitude and frequency of the periodic solutions with the corresponding analytical predictions, using a Fourier transform to determine the frequency of oscillation in the numerical case. Table \ref{tab:template} shows the results of such a comparison, for variable $Re$, using $D=1000$, $S=10$ and $\tau=0.1$ (which includes the ``reference case'' below). All results are accurate to within $1\%$ which clearly validates both the numerical scheme and the analytical calculations. Finally, fixing $Re=10$, $D=1000$ and $S=10$, we find that the solution exhibits a transition from periodic to quasi-periodic dynamo waves at $\tau=0.347$, which is compares very favourably to the analytic value of $\tau=0.348$ (see Table~\ref{tab:tcvaluesdpos}).

\begin{figure}
\begin{center}
\resizebox{\hsize}{!}{\includegraphics[angle=270]{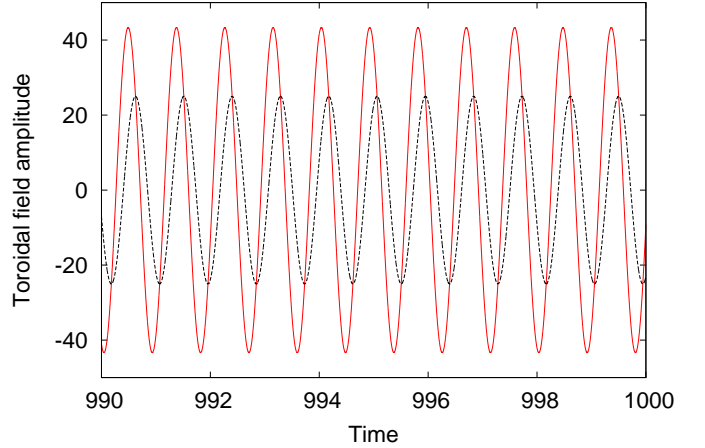}}
\caption{\label{fig:harmonic} The reference case for $\tau=0.2$: This shows the toroidal field $B$ (solid line) and the delayed toroidal field $Q$ (dashed line) as a function of time (which is expressed in dimensionless units).}
\end{center}
\end{figure}
\subsection{\label{rc}Results}

\begin{figure*}
\centering
\includegraphics[scale=0.35,angle=270]{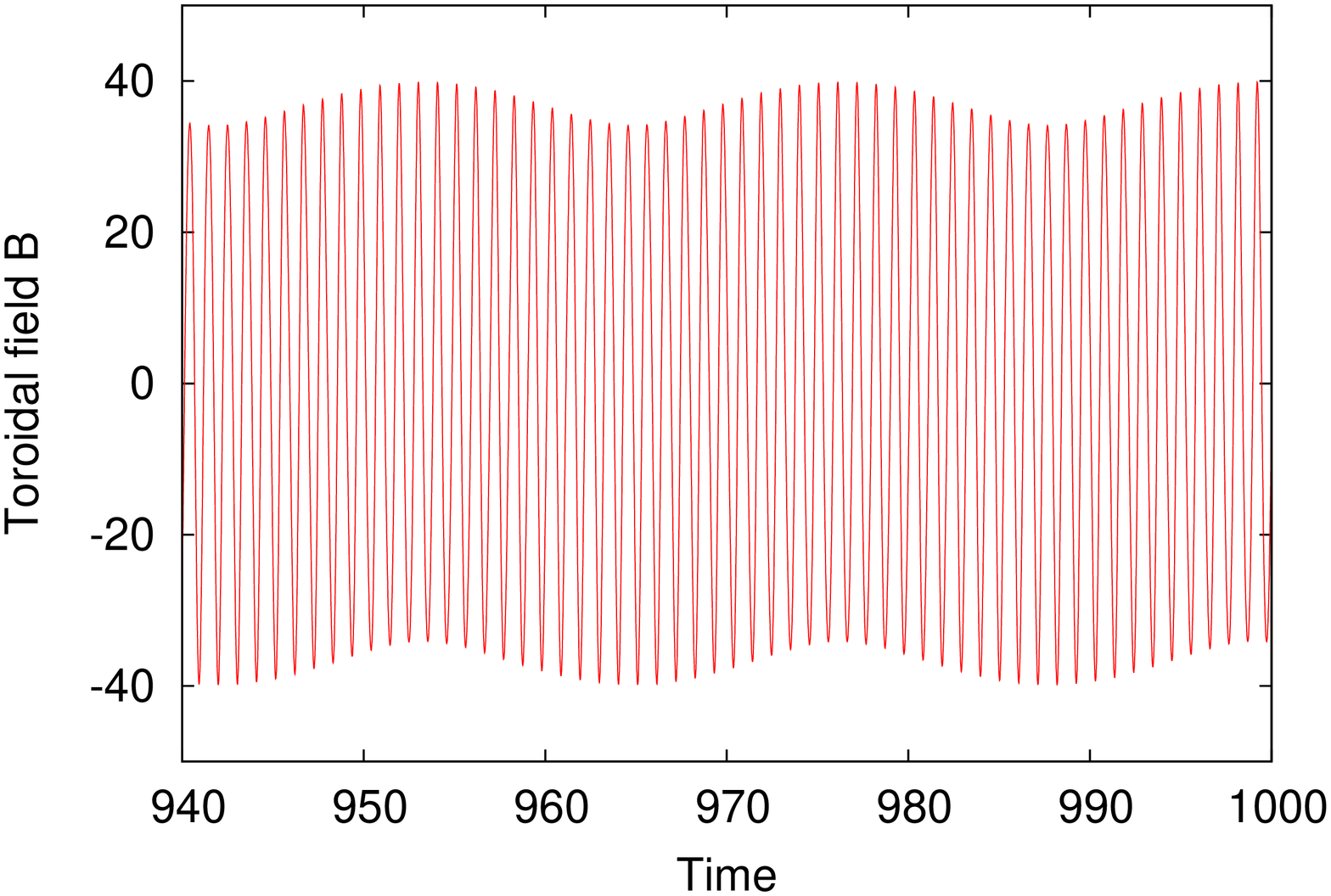}
\includegraphics[scale=0.35,angle=270]{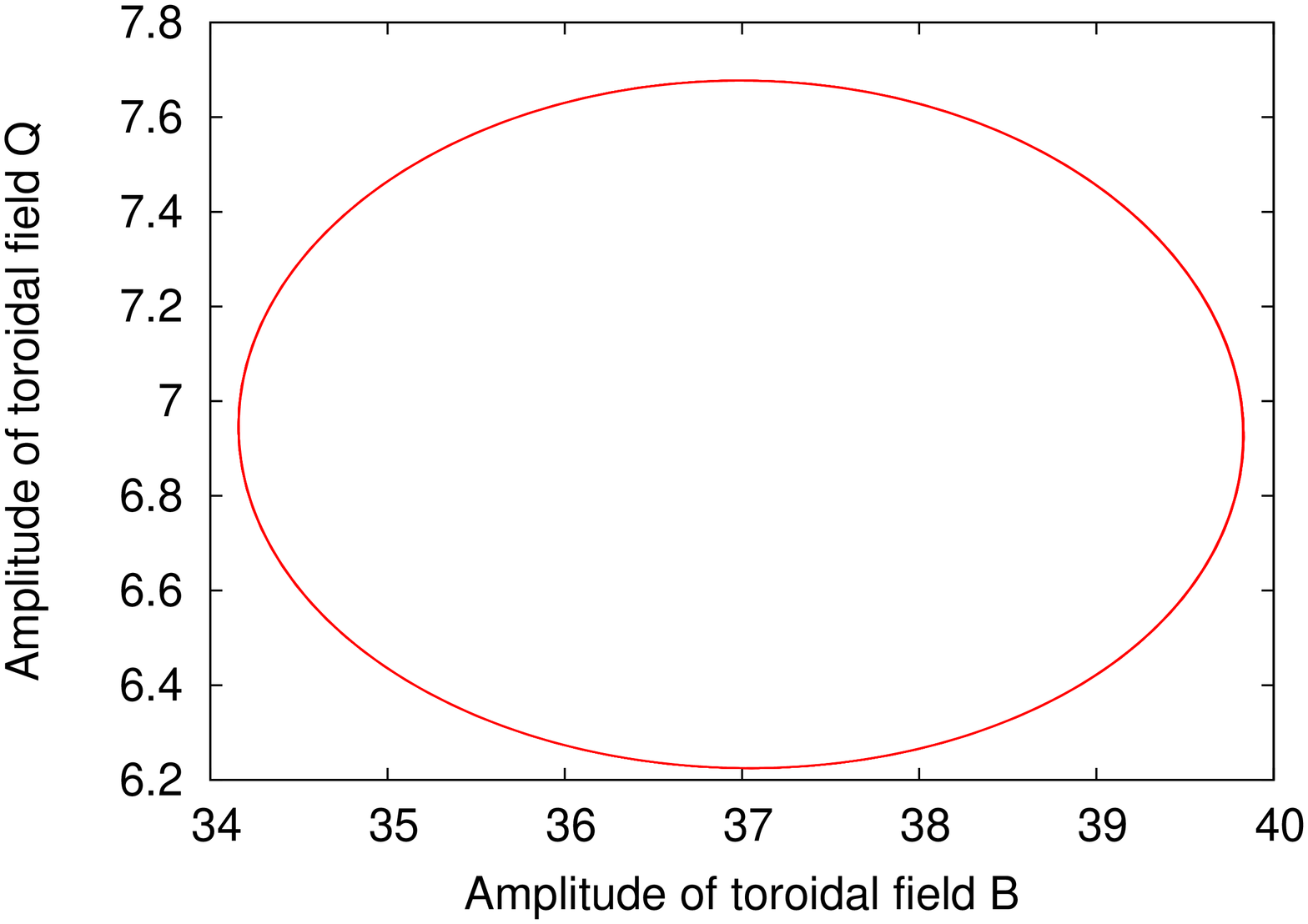}\\
\includegraphics[scale=0.35,angle=270]{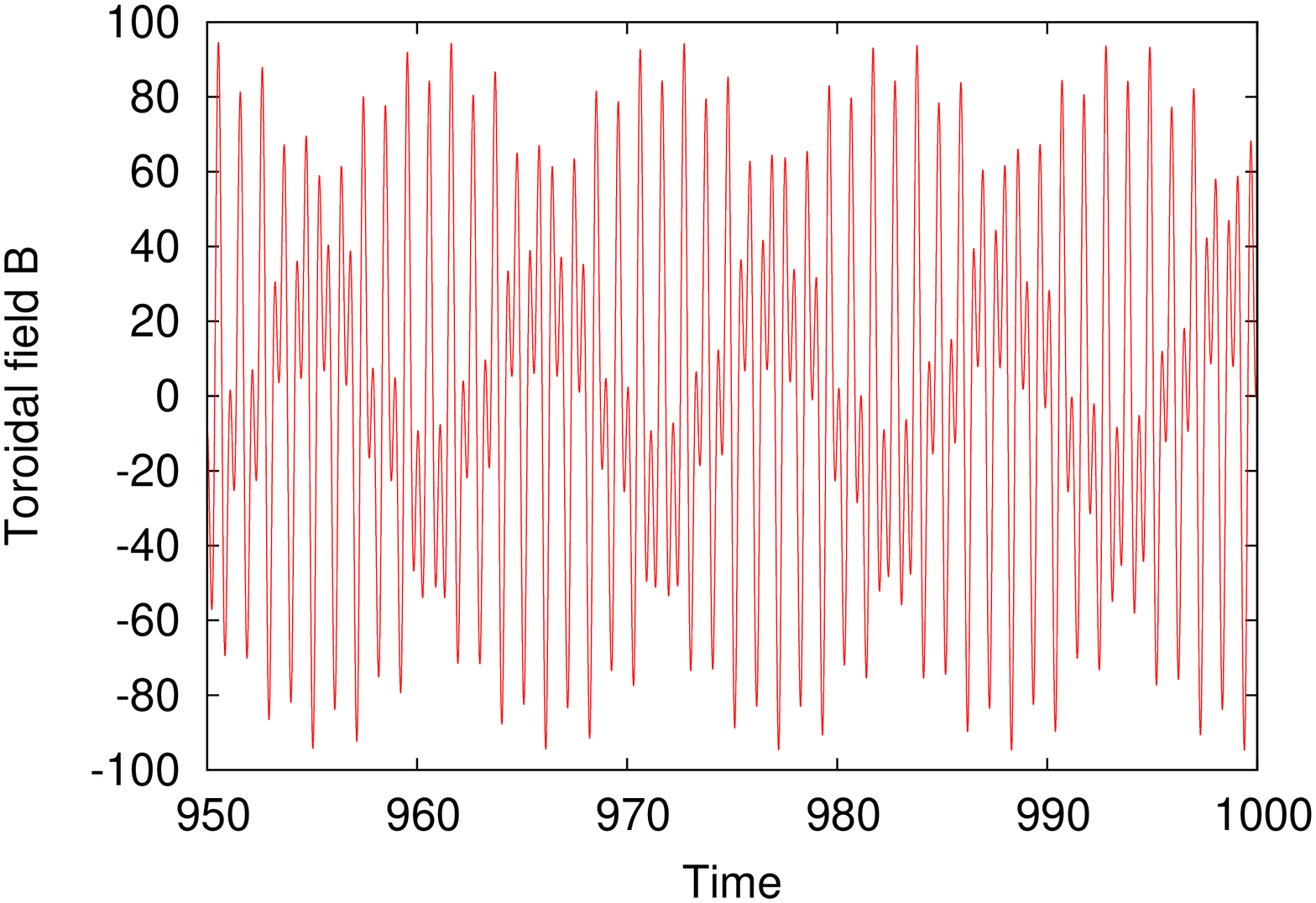}
\includegraphics[scale=0.35,angle=270]{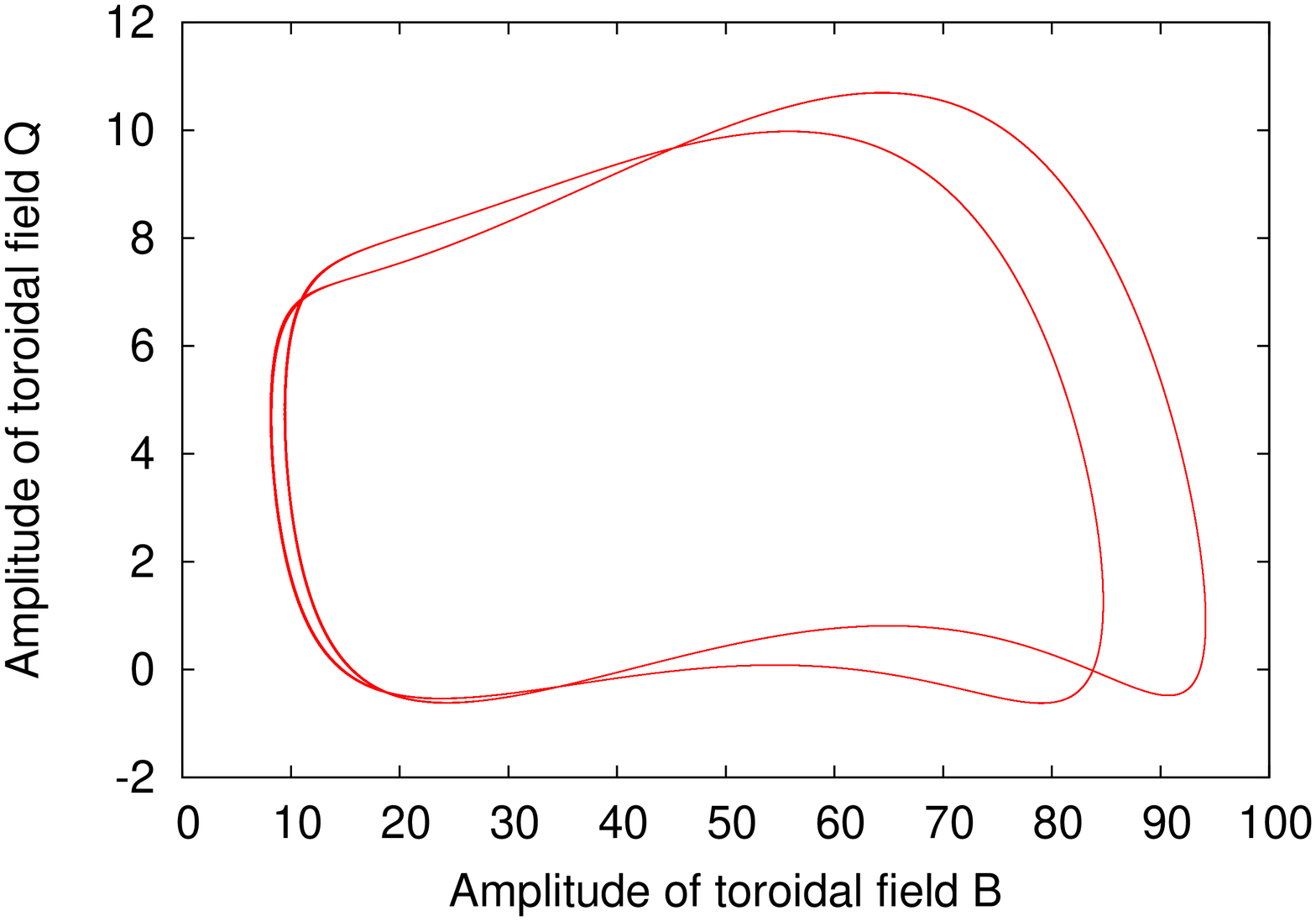}\\
\includegraphics[scale=0.35,angle=270]{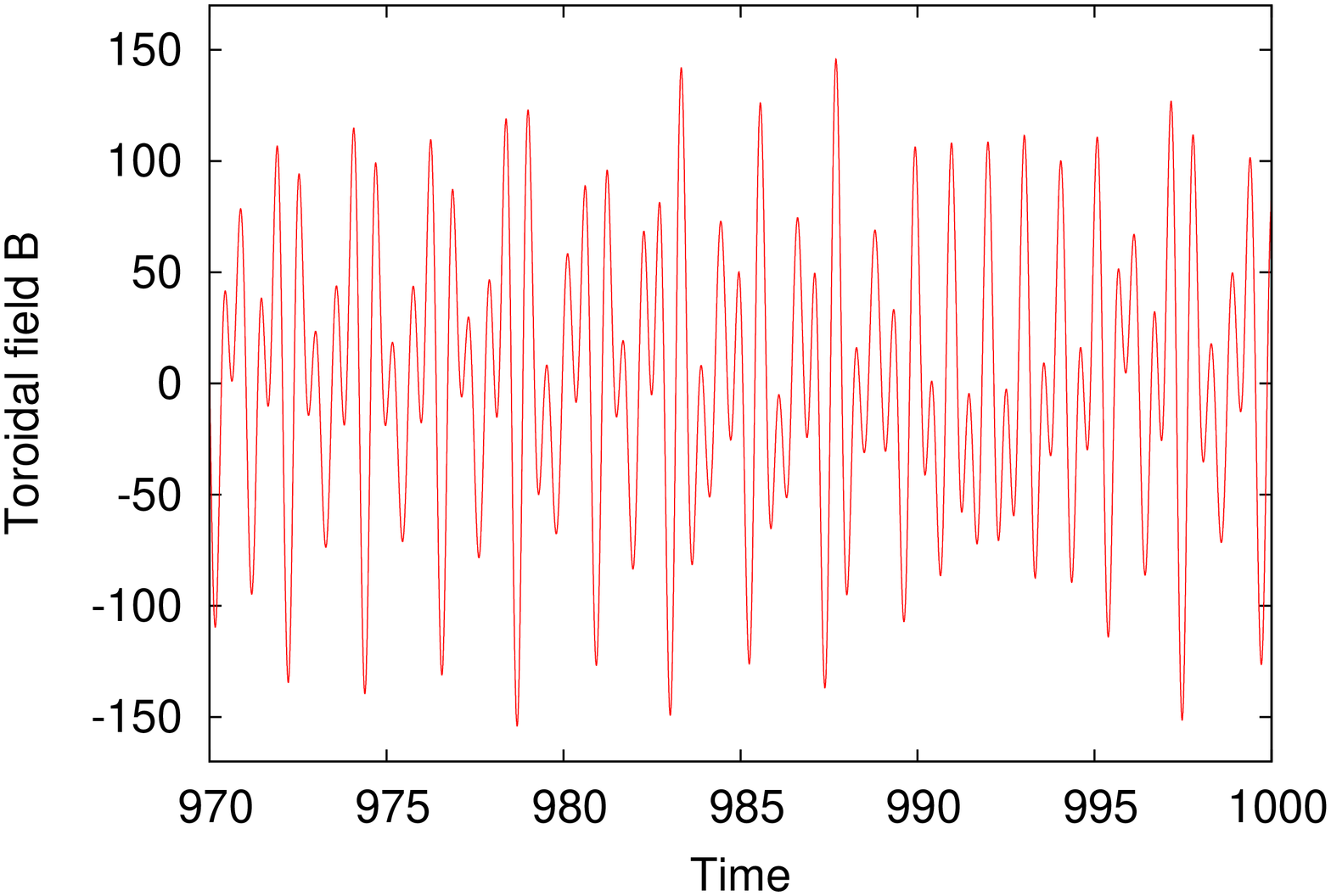}
\includegraphics[scale=0.35,angle=270]{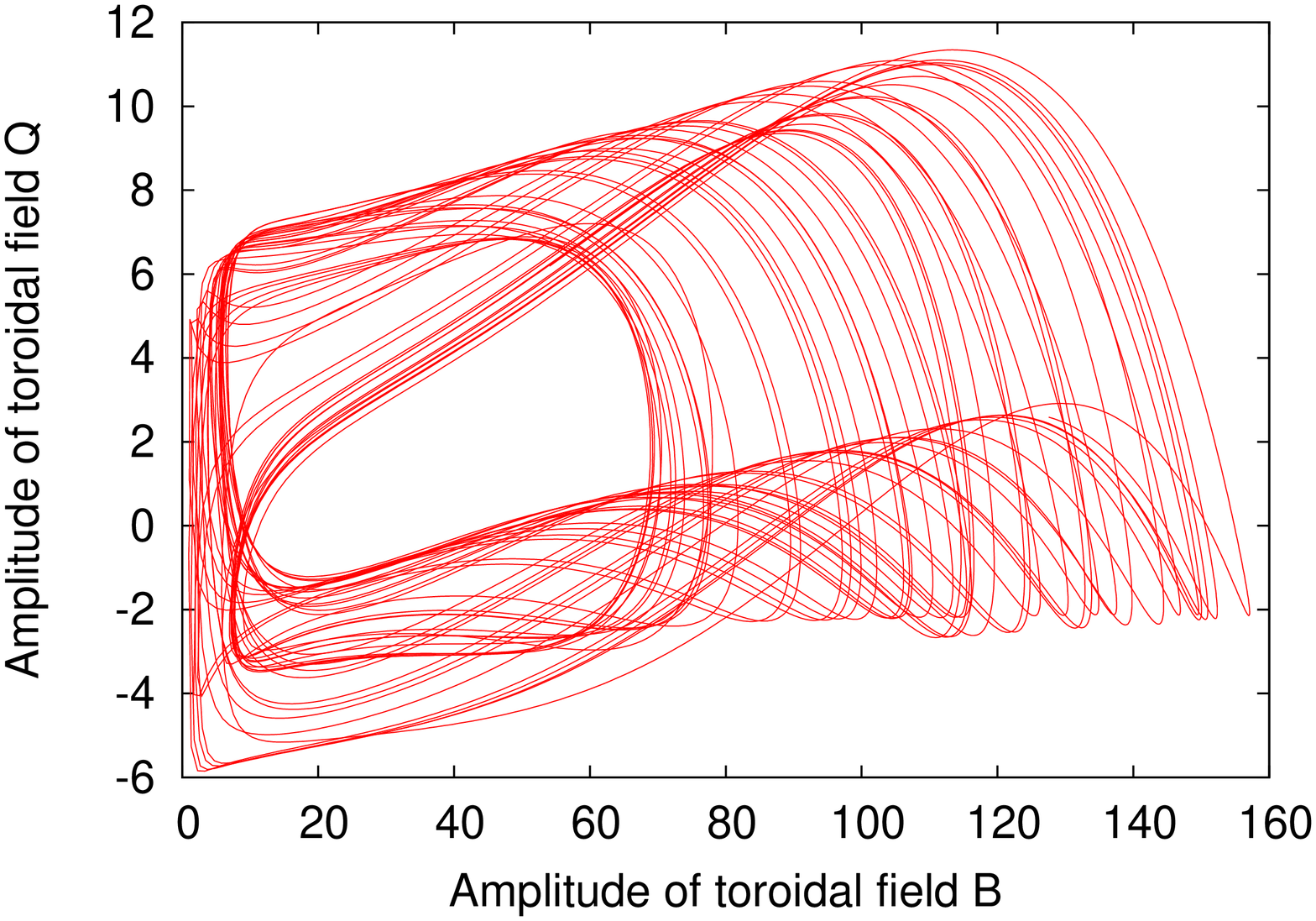}
     \caption{The reference case for $\tau=0.35$ (top), $\tau=0.61$ (middle) and $\tau=0.86$ (bottom). The plots on the left show the time-dependence of the toroidal field, whilst the plots on the right show the phase portraits of the amplitudes of $B(t)$ against $Q(t)$ (as derived from the $5^{\rm th}$-order system).}
     \label{fig:reference}
\end{figure*}

\begin{figure}
\begin{center}
\resizebox{\hsize}{!}{\includegraphics[angle=270]{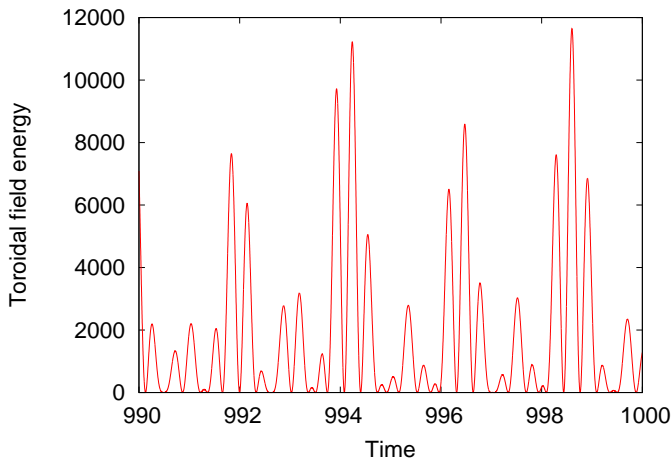}}
\caption{\label{fig:bsq86}The reference solution for $\tau=0.86$. A plot of $B^2$ against time.}
\end{center}
\end{figure}

Initially, the parameters are chosen such that $D=1000$, $Re=10$, $S=10$ and $\tau$ is varied (we will refer to this as the ``reference case''). Figure \ref{fig:harmonic} shows that a periodic solution can be found provided that $\tau$ is sufficiently small. Both $B(t)$ and $Q(t)$ oscillate with constant amplitude although $Q(t)$ has a smaller amplitude of oscillation and, as expected, lags behind $B(t)$. The effects of increasing the value of $\tau$ are shown in Figure \ref{fig:reference}. As $\tau$ is increased through the threshold value of $\tau=0.347$, the lag between $B(t)$ and $Q(t)$ increases to such an extent that we see a transition to a quasi-periodic state. Further increases in $\tau$ lead to further transitions, from multiply periodic to chaotically modulated states. Figure \ref{fig:bsq86} shows the time evolution of the toroidal field energy $B^2$ for $\tau=0.86$, at which point the solution is chaotically modulated. It is clear that there are several phases of significantly reduced magnetic activity, and it is tempting to compare these to grand minima. The extent to which this behaviour is ``solar-like'' is a matter of some debate -- this is, after all, a highly idealised model. Nevertheless, it is encouraging that this simple model, with competing $\alpha$-effects, is capable of producing highly modulated dynamo waves when the time delay is large.  


\par
\begin{figure}
\begin{center}
\resizebox{\hsize}{!}{\includegraphics[angle=270]{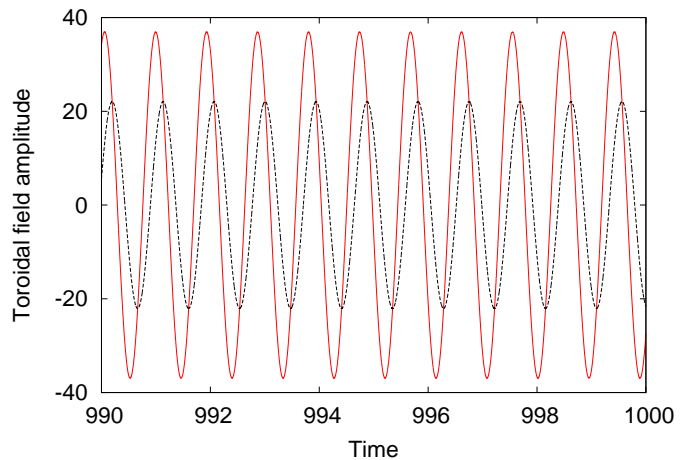}}
\resizebox{\hsize}{!}{\includegraphics[scale=0.53,angle=270]{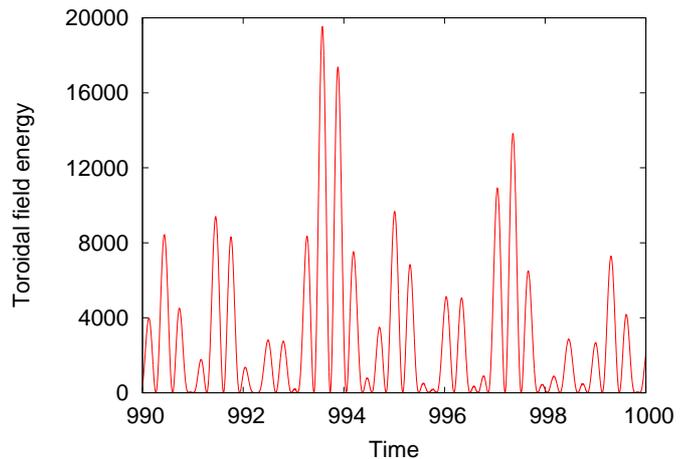}}
\caption{\label{fig:dnegharmonic} The effects of increasing $\tau$ for $D=-1000$,  $Re=10$, $S=-10$. The upper plot shows the time evolution of the toroidal field $B$ (solid line) and the delayed toroidal field $Q$ (dashed line) for a periodic solution at $\tau=0.2$. The lower plot shows the toroidal field energy, $B^2$, as a function of time for $\tau=1.08$. }
\end{center}
\end{figure}

 As indicated by the results in Table~\ref{tab:tcvaluesdpos}, the analysis of the stability of the periodic solution indicates that it is not possible to find a transition to a quasi-periodic solution for negative values of $S$, when $D$ is positive. This tendency for the periodic state to be stable (for $S\le 0$) regardless of the value of $\tau$ has been confirmed numerically. However, for positive values of $S$ it always appears to be possible to find a transition to quasi-periodic solutions, provided that the Reynolds number is non-zero, and these transitions are consistent with those predicted in Table~\ref{tab:tcvaluesdpos}. Once quasi-periodic solutions have been found it is usually possible to find chaotically-modulated states for sufficiently large values of the time delay. For negative values of the dynamo number, the results are again consistent with those predicted analytically. No modulation is found for positive $S$ or for $Re=0$. For negative $S$ and positive $Re$, it is possible to find quasi-periodic and chaotically modulated solutions as the time delay is increased. One such solution is illustrated in Figure~\ref{fig:dnegharmonic}. As in Figure~\ref{fig:bsq86}, it should again be noted that the chaotically modulated solution that is illustrated in the lower part of Figure~\ref{fig:dnegharmonic} is characterised by phases of significantly reduced magnetic activity. 

\section{\label{sec:pdesim} Solving the PDE system}

Although the results from the local model are promising, it is important to verify that they are not crucially dependent upon the simplifying assumptions that have been made when deriving the model. In this section, we return to the original model of partial differential equations, as defined by Equations~(\ref{PDEfinala}) -- (\ref{PDEfinalq}). In dimensionless units, we assume that $0 \le x \le \pi/2$ (recalling that we interpret $x$ as being analogous to the co-latitude on a spherical surface), imposing the boundary conditions that $A=B=Q=0$ at $x=0$ (the ``North pole'') and $B=Q=\partial A/\partial x=0$ at $x=\pi/2$ (the ``Equator''). These boundary conditions correspond to the assumption that the global magnetic field has dipolar symmetry. Having neglected the effects of curvature, and having assumed constant $\alpha_0$, $v_0$ and $\Omega_0$, we should stress again that this should still be regarded as an illustrative model. Nevertheless, it contains the key physical ingredient of two competing $\alpha$-effects with a surface $\alpha$-effect contribution that depends upon a time-delayed toroidal field. In order to obtain dynamo waves that propagate towards the Equator, we focus primarily upon the $D<0$ parameter regime (which would correspond to a negative deep-seated $\alpha$-effect in the northern hemisphere). We solve the governing equations numerically, approximating derivatives using second-order finite differences. A 4th-order Runge-Kutta scheme is again used to time-step the governing equations.

\par Given that we are investigating the $D<0$ regime, the local model suggests that we should be able to find modulation for  negative values of $S$. However, in this region of parameter space there is an overwhelming tendency for steady modes to be preferred at onset (recall that wavelike solutions were assumed when the local model was derived). It is well known that steady and oscillatory modes can bifurcate from the trivial state at similar values of $D$ in global $\alpha\Omega$ dynamos \citep[see, for example,][]{JENWEI}, so this behaviour is not entirely unsurprising. However, it is almost certainly rather model specific -- experimentation with the inclusion of different nonlinear quenching mechanisms suggests that it is possible to obtain oscillatory solutions in these parameter regimes. Furthermore, oscillatory solutions can be found for positive dynamo numbers and therefore, despite some differences, the results from the local model should not be discarded.

\par
\begin{figure}
\begin{center}
\subfigure{\includegraphics[scale=0.35,angle=270]{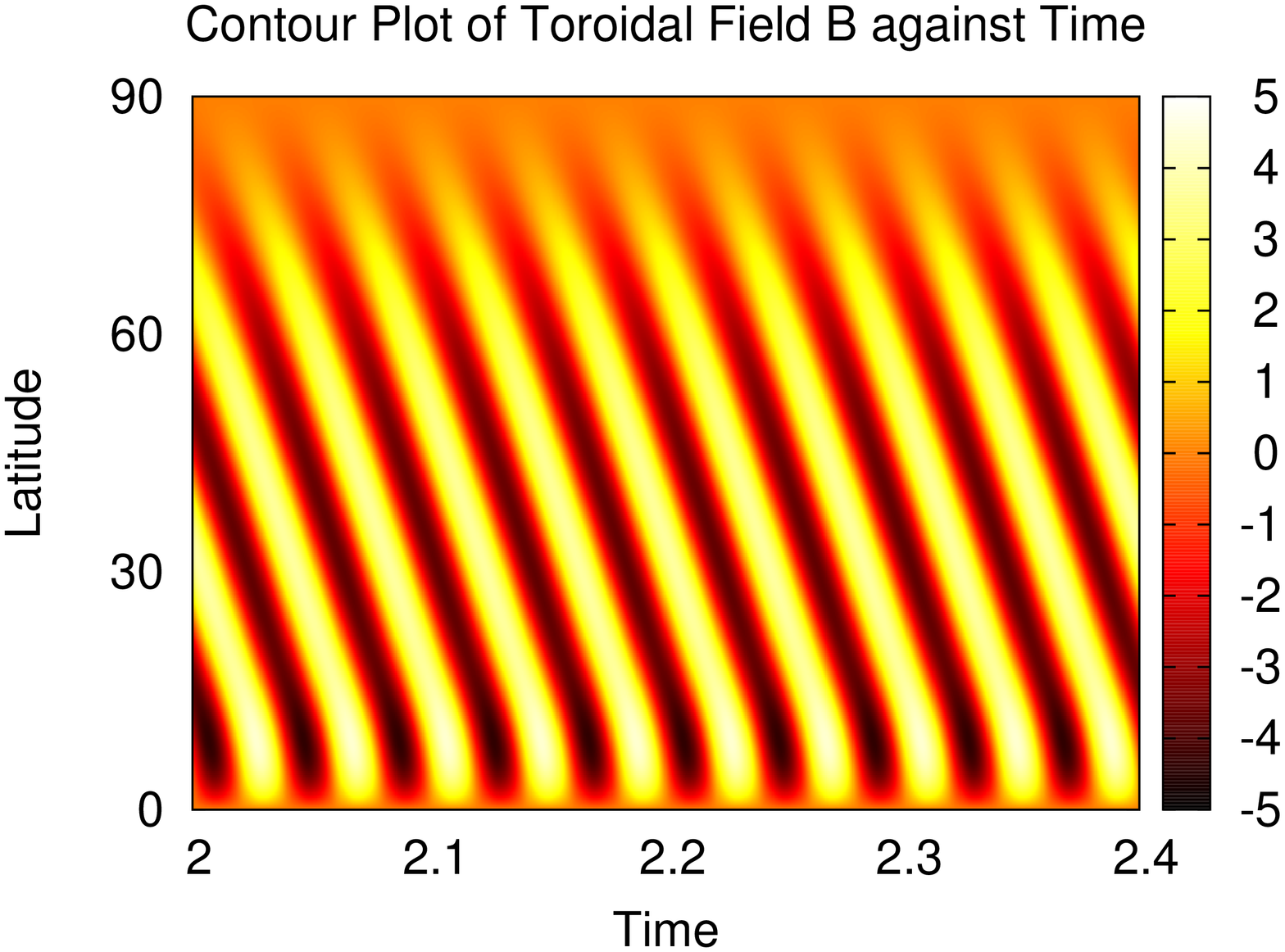}}
\subfigure{\includegraphics[scale=0.35,angle=270]{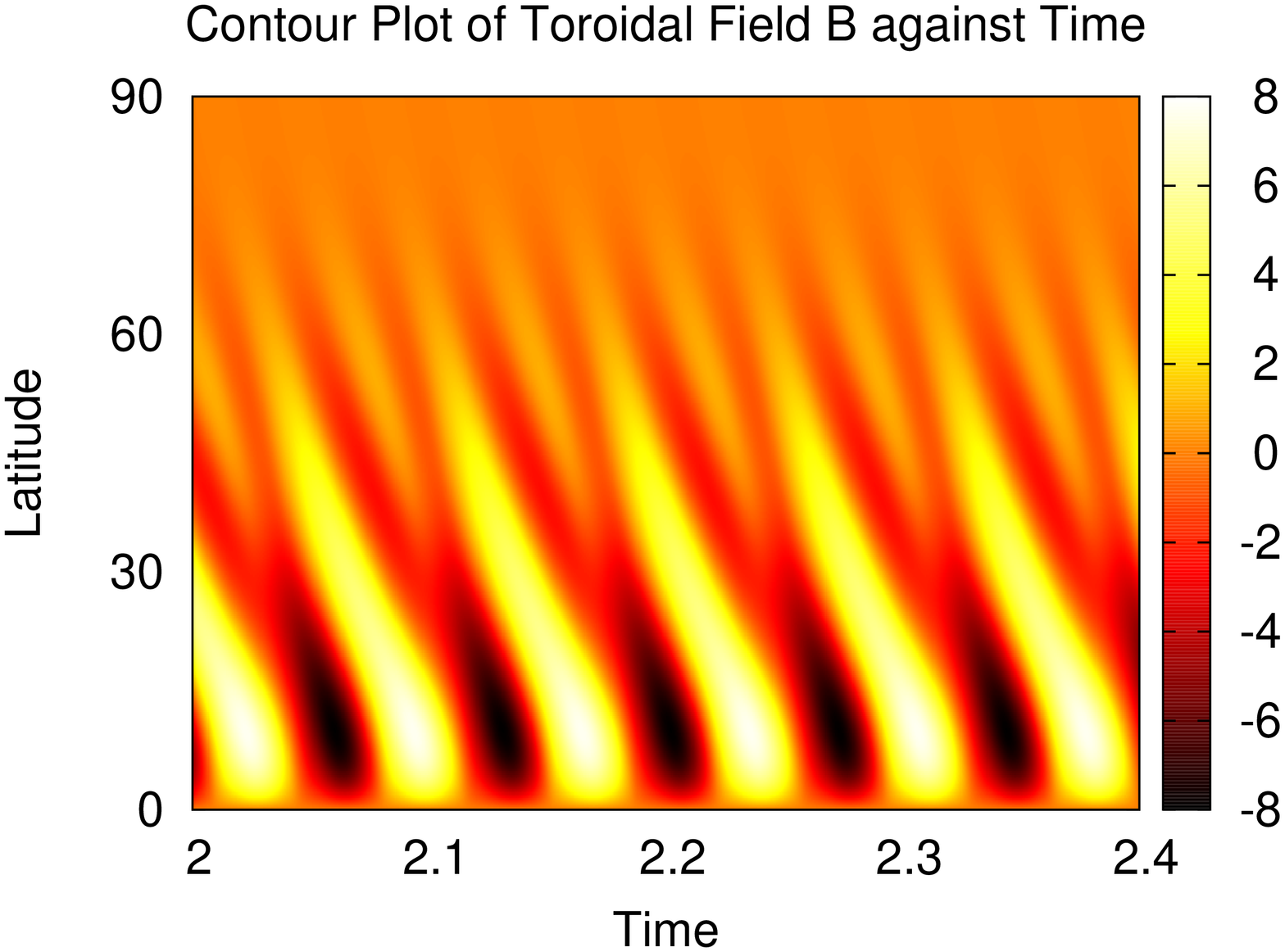}}
\subfigure{\includegraphics[scale=0.35,angle=270]{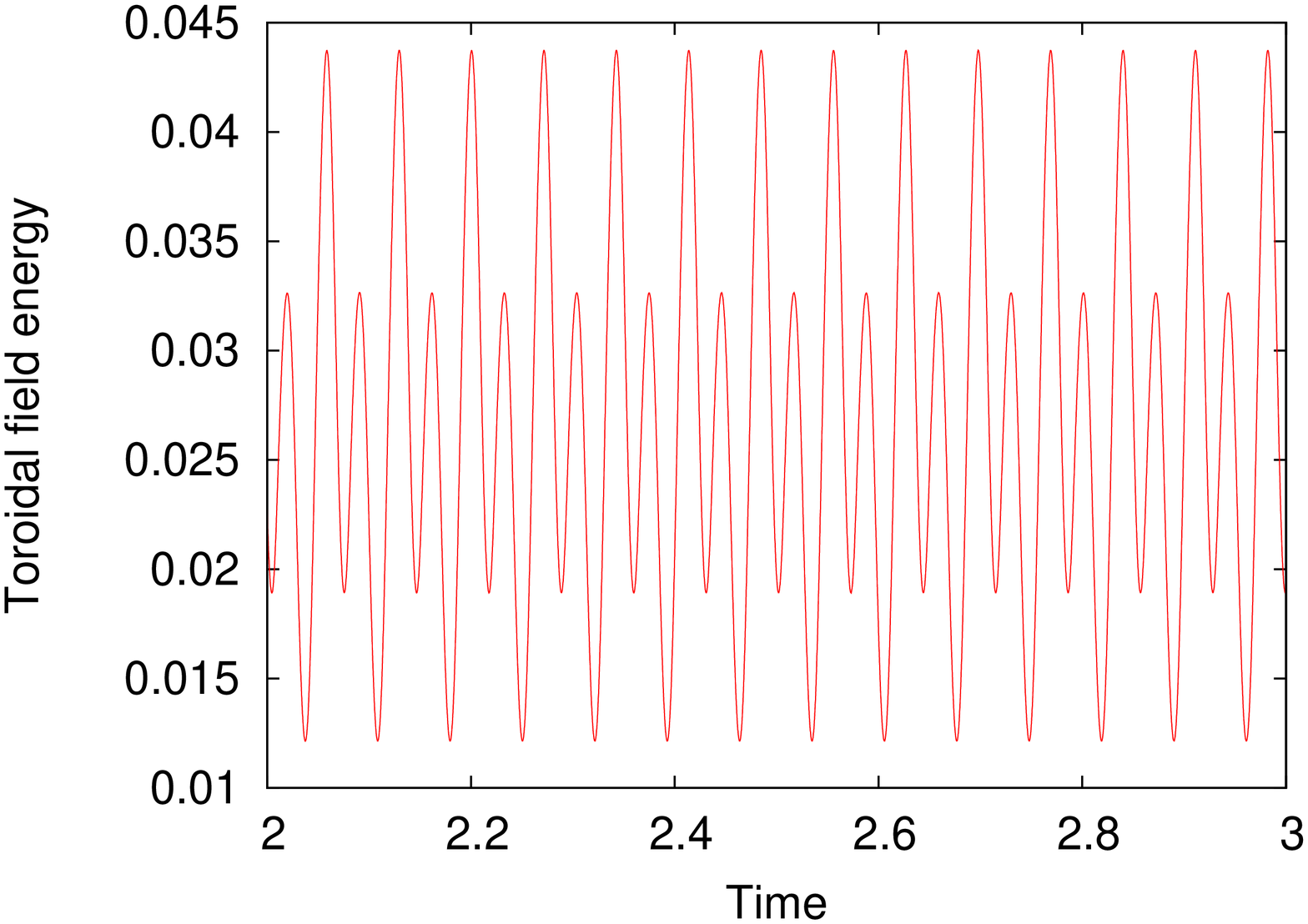}}
\caption{Dynamo solutions from the full PDE system ($D=-6000$, $S=1$ and $Re=10$). Top: contours of toroidal field as a function of latitude and time (a latitude of 90$^\circ$ corresponds to the pole, 0$^\circ$ to the equator) for $\tau=0.01$. Middle: as above, but for $\tau=0.05$. Bottom: a plot of the energy in the toroidal field as a function of time for the quasi-periodic solution that is obtained for $\tau=0.05$.}
\label{fig:contlow}
\end{center}
\end{figure}

\begin{figure}
\begin{center}
\subfigure{\includegraphics[scale=0.35,angle=270]{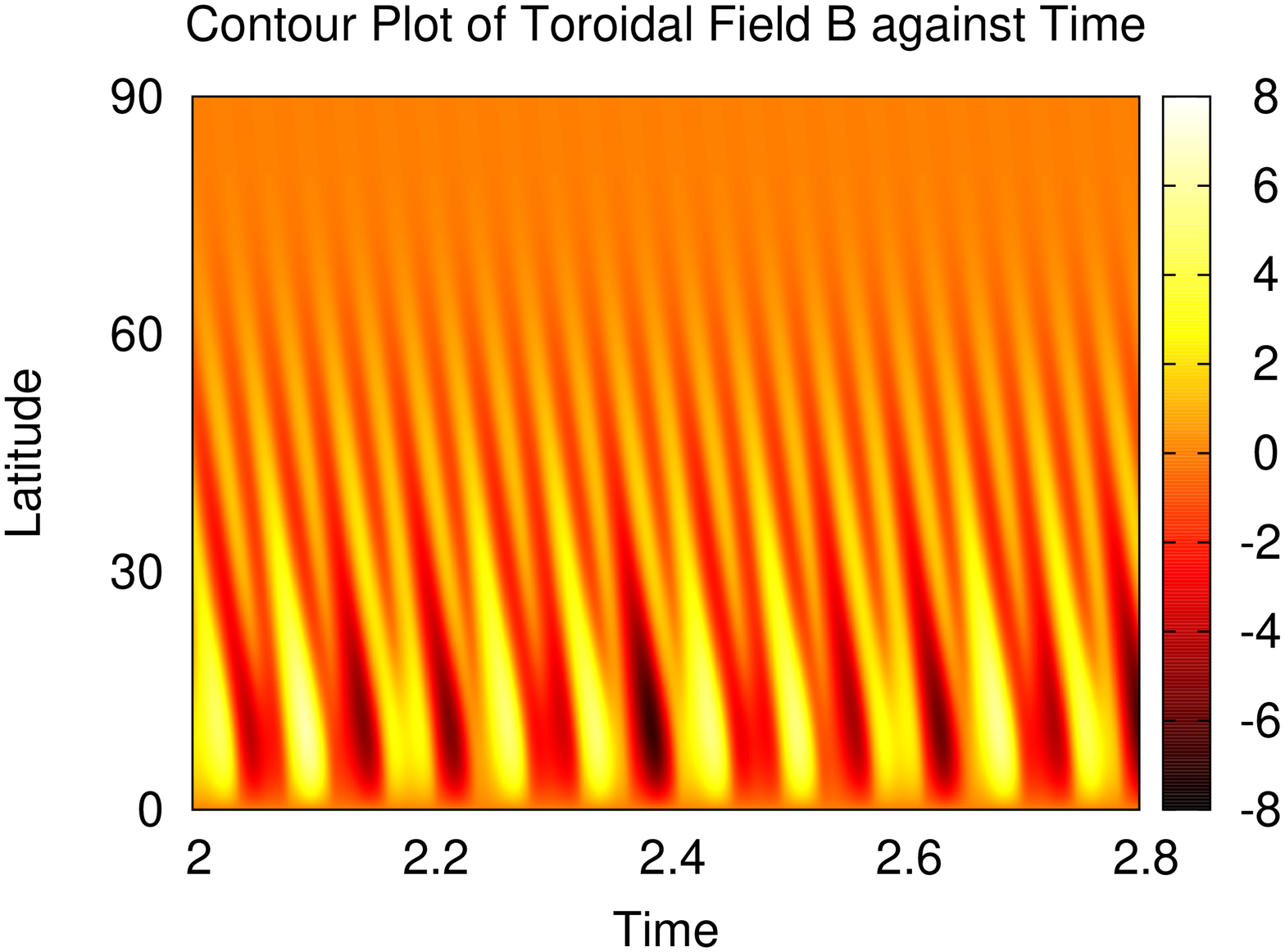}}
\subfigure{\includegraphics[scale=0.35,angle=270]{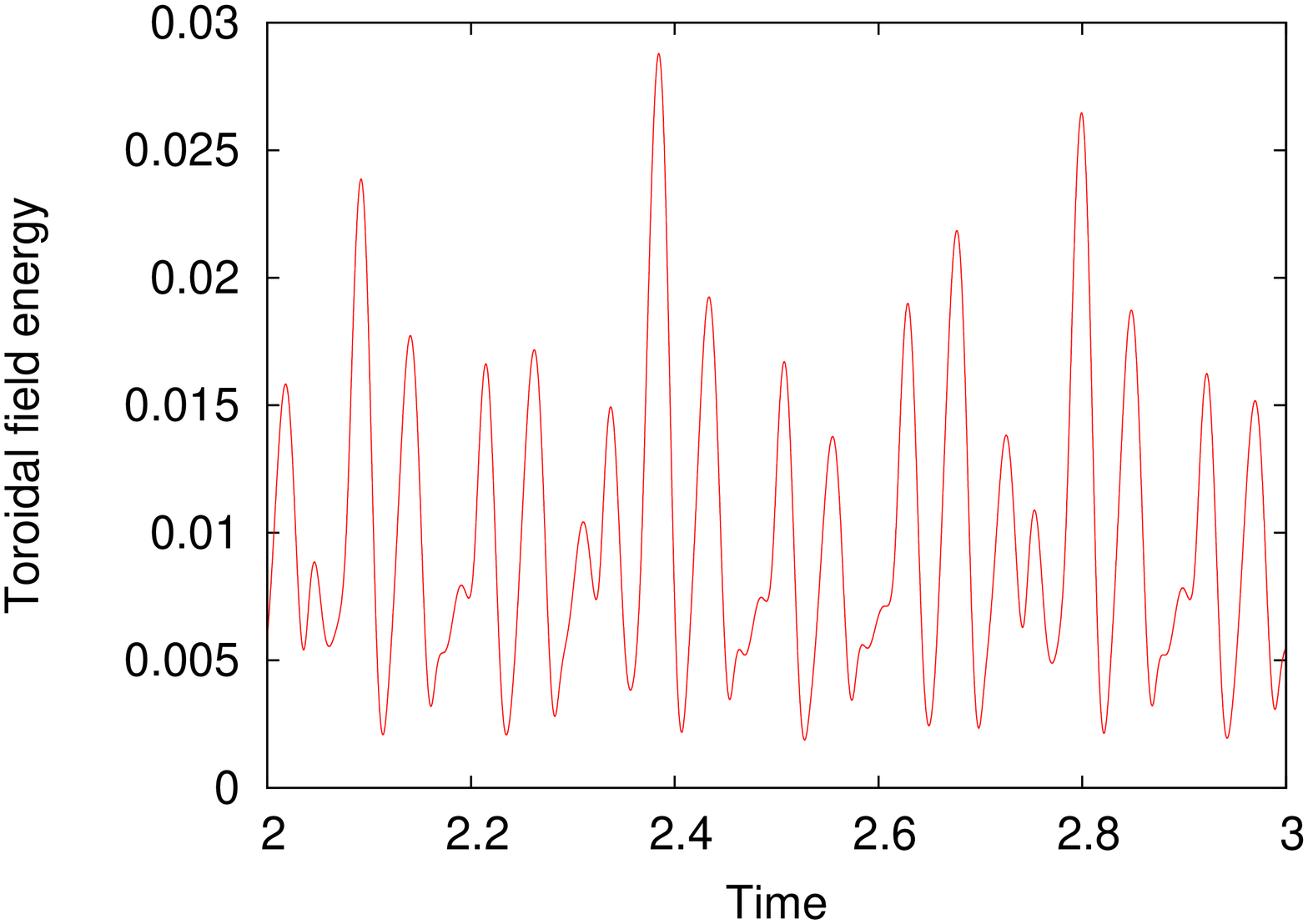}}
\caption{Chaotically-modulated solution from the full PDE system ($D=-6000$, $S=1$, $Re=10$ and $\tau=0.09$). Top: contours of toroidal field as a function of latitude and time. Bottom: a plot of the energy in the toroidal field as a function of time.} \label{fig:conthigh}
\end{center}
\end{figure}

 In fact, in the case of this global model, interesting solutions can be found for negative values of $D$ and positive values of $S$. This is illustrated by Figure~\ref{fig:contlow} which shows solutions for $D=-6000$, $S=1$ and $Re=10$. A periodic solution can be found at $\tau=0.01$. This is characterised by an oscillatory magnetic field which propagates towards the Equator (note that these contour plots have been plotted as a function of latitude and time, for ease of comparison with observations). Increasing the time-delay leads to a transition to a quasi-periodic solution. Further increases in $\tau$ eventually lead to chaotically modulated oscillations (as illustrated in Figure~\ref{fig:conthigh}). This solution is rather ``solar-like'' in many respects, with the dynamo confined to low latitudes, and with strong variations in the amplitudes of successive cycles. Furthermore, the modulation is characterised by periods of reduced magnetic activity. So although the modulation due to these competing $\alpha$-effects was not in the expected parameter regime, it is clearly a robust feature of this system. 

\section{Conclusions}

In this paper, we have investigated the properties of an illustrative mean-field dynamo model which includes two competing $\alpha$-effects. The first of these is the standard deep-seated $\alpha$-effect, the second is due to a surface $\alpha$-effect (of Babcock-Leighton type). Following the approach described by \citet{JOUVE}, who did not consider competing $\alpha$-effects, the contribution from the surface $\alpha$-effect was modelled by assuming that it depends upon a time-delayed toroidal field (with a constant parameterised time delay $\tau$). Two different approaches were applied to this model. Initially, a local approximation was made to reduce the governing equations to a system of coupled ordinary differential equations. A linearised version of these equations was used to determine the dependence of the critical dynamo number upon $S$ (the magnitude of the surface $\alpha$-effect) and $\tau$. Generally, the larger the magnitude of $S$, the easier it becomes to excite the dynamo. However, there are some regions of parameter space in which the two competing $\alpha$-effects appear to impede each other, thus inhibiting the dynamo. Moving beyond linear theory, it was found that there are significant regions of parameter space in which the periodic solution becomes unstable with increasing $\tau$, leading to quasi-periodicity. This was verified numerically, where further increases in the time delay were shown to produce chaotically modulated states with phases of significantly reduced activity. The full PDE model was then investigated. Although modulation was found, this occurs in a different parameter regime to that predicted by the ODE model. This discrepancy could be model specific, although we expected to see some differences between the two models due to the fact that significant simplifications were made when deriving the set of coupled ODEs. Nevertheless, it was possible to find chaotically modulated solutions in the PDE model, and these solutions exhibit certain features that are (at least qualitatively) ``solar-like''.

\par There are many possible areas of future work. In particular, more could be done to explore the robustness of the PDE model to variations in the boundary conditions and the non-linear quenching mechanisms. As has already been mentioned, preliminary calculations suggest that the adoption of different nonlinearities may make a very significant difference to the behaviour of the model. It may also be possible to improve the existing model by refining the way in which the time delay is implemented -- the current approach is simple and effective, but is derived by truncating a Taylor series expansion at lowest order. Retaining higher order terms may make a difference to the behaviour of the model. Moving beyond the one-dimensional Cartesian system, it would be natural to explore a two-dimensional version of this model in (axisymmetric) spherical geometry. This would open up the possibility of including a more realistic flow geometry (in both the meridional and azimuthal directions) as well as spatially dependent mean-field coefficients. Although this would still be within the framework of mean-field theory, a more realistic model would enable more detailed comparisons to be made between our results and the solar dynamo. 


\bibliographystyle{aa}
\bibliography{paper}
\end{document}